\documentclass[aps,prl,twocolumn,groupedaddress,showpacs]{revtex4}

\usepackage{graphicx}

\begin{document}

\title{Dynamical resurrection of the visibility in a Mach-Zehnder
interferometer}


\author{A.V.\ Lebedev and G.\ Blatter}

\affiliation{Theoretische Physik, Wolfgang-Pauli-Strasse 27, ETH Zurich,
CH-8093 Z\"urich, Switzerland}

\date{\today}

\begin{abstract}

We study a single-electron pulse injected into the chiral edge-state
of a quantum Hall device and subject to a capacitive Coulomb
interaction.  We find that the scattered multi-particle state
remains unentangled and hence can be created itself by a suitable
classical voltage-pulse $V(t)$. The application of the inverse pulse
$-V(-t)$ corrects for the shake-up due to the interaction and
resurrects the original injected wave packet.  We suggest an
experiment with an asymmetric Mach-Zehnder interferometer where the
application of such pulses manifests itself in an improved
visibility.

\end{abstract}


\maketitle

On demand single-electron sources are an essential building block on the road
to a mesoscopic solid state implementation of quantum computing.
Single-particle wave packets can be generated with the help of suitable
voltage pulses \cite{ivanov97} and a first experimental realization of such a
source has been recently achieved \cite{Feve07} in a quantum Hall setup.
Contrary to their photonic counterparts, such single-electron states are prone
to decoherence due to the interaction with the underlying Fermi-sea
\cite{degiovanni09}.  Here, we study the influence of a capacitive Coulomb
interaction on a single-electron wave-packet injected into the chiral edge
state of a quantum Hall device. Due to the interaction, the injected particle
transfers energy to the Fermi sea, leading to the shake-up of electron-hole
pairs. Analyzing the resulting scattered state, we find that it corresponds to
a simple Slater determinant; the underlying product nature of the resulting
multi-particle state allows one to undo the decoherence by applying a suitable
local voltage-pulse.

The resurrection of decohered single-particle wave packets has numerous
potential applications; here, we suggest to test this prediction in a
Mach-Zehnder interferometer implemented in a quantum Hall setup. Electronic
decoherence has become apparent in such devices \cite{yang03} through the
observation of a non-trivial decay of the visibility \cite{neder06,exper} with
increasing bias voltage and a satisfactory explanation could be obtained
\cite{sukhorukov0708} via accounting for strong Coulomb interaction between
edge states.  The experiments~\cite{yang03,neder06,exper} have been performed
with a finite bias voltage where electrons are stochastically injected into
the system.  We suggest to use an asymmetric setup operating in the $\nu=1$
quantum Hall regime, where the decoherence is introduced in a controlled
manner through a capacitive coupling of one arm to a metallic gate.
Applying suitable voltage pulses to the scattered wave function behind the
interaction region, the visibility of the interference pattern can be improved
considerably though not perfectly, a consequence of our ignorance regarding
the path which the electron has taken in traversing the device.

In the following, we study a one-dimensional ballistic conductor with chiral
spinless electrons propagating to the right. The capacitive Coulomb
interaction is described \cite{Pretre96} by the Hamiltonian
$\hat{H}_\mathrm{int} =\hbar\omega_{\rm \scriptscriptstyle C}\, \hat{N}^2/2$,
where $\hat{N} = \int dx\, g(x) :\!\!\hat\Psi^\dagger(x) \hat\Psi(x)\!\!:$ is
the effective number of excess electrons within the interaction region defined
by the coordinate kernel $g(x)$, $\hbar\omega_{\rm \scriptscriptstyle C}$ is
the Coulomb energy, and $:\!\hat{A}\!: \>\equiv \hat{A} - \langle
\Phi_{\rm\scriptscriptstyle F}| \hat{A} |\Phi_{\rm\scriptscriptstyle
F}\rangle$ denotes normal ordering with respect to the Fermi sea
$|\Phi_{\rm\scriptscriptstyle F}\rangle$. Within our setup, we ignore any
dissipative coupling to the environment \cite{degiovanni09} and study the
effect of the interaction on the structure of the scattered many-body wave
function.

Adopting a Luttinger Liquid description \cite{Delft_Sch_98}, we introduce the
chiral bosonic field $\hat\theta(x) = -\sum_{k>0} (\hat{b}_k e^{ikx}$
$+\hat{b}^\dagger_k e^{-ikx})/\sqrt{k}$ obeying the commutation relation
$[\hat\theta(x), \theta(x^\prime)] = i\pi\,\mbox{sgn}(x-x^\prime)$, where
$\hat{b}_k^\dagger$, $\hat{b}_k$ are bosonic creation and annihilation
operators, $[\hat{b}_k, \hat{b}_{k^\prime}^\dagger] = \delta_{kk^\prime}$. The
electron field operator $\hat\Psi(x)$ and the electronic density fluctuations
$\hat\rho(x) =\> :\!\hat\Psi^\dagger(x) \hat\Psi(x)\!:$ can be expressed via
the field $\hat\theta(x)$ as $\hat\Psi(x) = (\hat{F}/\sqrt{2\pi\delta})\,
\exp[-i\hat\theta(x)]$ and $\hat\rho(x) = \partial_x \hat\theta(x)/2\pi$,
where $\hat{F}$ is the Klein factor acting as a Fermion-number ladder operator
and $\delta$ is an ultraviolet cutoff.

The interaction Hamiltonian is quadratic in the bosonic field $\hat\rho(x)$,
allowing for an exact solution of the equations of motion for
$\hat\theta(x,t)$, see Ref.~\cite{sukhorukov0708}. Here, we choose a different
approach and apply a Hubbard-Stratonovich transformation with an auxiliary
real field $z(t)$ in order to express the evolution operator
$\hat{S}(-\infty,\infty)$ as an exponential linear in $\hat{N}(t)$,
\begin{eqnarray}
      \hat{S} &=& \hat{T}_+
      \exp\Bigl[-\frac{i\omega_{\rm \scriptscriptstyle C}}2\,\int
      dt\,\hat{N}^2(t)\Bigr]
      \\
      &=& \int D[z]\, \hat{T}_+
      \exp\Bigl[ i\omega_{\rm \scriptscriptstyle C} \int dt \bigl(
      z^2(t) - z(t)\hat{N}(t) \bigr) \Bigr],
      \nonumber
\end{eqnarray}
with $\hat{T}_+$ the usual (forward) time ordering operator. Below, we will be
interested in a non-stationary situation and thus define the evolution
operator $\hat{S}_K$ as a time ordered exponent along the Keldysh contour,
where $z(t)$ assumes different values $z_\pm(t)$ on the upper and lower branch
of the contour \cite{norm}.  Correlation functions are defined as $\langle
\hat{T}_K \{ \hat{S}_K \hat{A}(t_1^{\mu_1}) \hat{B}(t_2^{\mu_2})
\dots\}\rangle$, where the Keldysh indices $\mu_i\in\{\pm\}$ specify the
branch of the Keldysh contour for the corresponding time instant $t_i$ and the
average is taken over the Fermi sea $|\Phi_{\rm\scriptscriptstyle F}\rangle$.

We consider the scattering problem where an electron wave packet $f(x)$ is
injected at the left above the Fermi sea, $|f\rangle = \int dx\, f(x-x_0)
\hat\Psi^\dagger(x,t_0) |\Phi_{\rm\scriptscriptstyle F}\rangle$, far from the
interaction region (we assume $t_0\rightarrow -\infty$ and
$x_0=v_{\rm\scriptscriptstyle F} t_0$, $v_{\rm\scriptscriptstyle F}$ the Fermi
velocity). At time $t=0$, the wave-packet reaches the interaction region where
additional electron-hole pairs can be excited.  Finally, as $t\rightarrow
+\infty$, the resulting many-particle scattering state $|\tilde{f}\rangle =
\int dx\, f(x-x_0)\, \hat{S}\,\hat\Psi^\dagger(x,t_0) |\Phi_{\rm
\scriptscriptstyle F}\rangle = \hat{S}|f\rangle$ involves the
excess electron dressed by a cloud of electron-hole excitations propagating to
the right.

Our focus is on the complexity of the scattered state $|\tilde{f}\rangle$. The
simplest fermionic many-particle state is a Slater determinant, i.e.,
an anti-symmetrized ($\tilde{\cal A}$) product state (obviously, our initial
state $|f\rangle$ is of this type). In fact, Slater determinants ${\cal A} (
|\phi_1\rangle \otimes |\phi_2\rangle \otimes \dots \otimes |\phi_N\rangle )$
correspond to non-entangled $N$-partite states which can be created trivially
by collecting fermionic particles in quantum states $\{ |\phi_i \rangle
\}_{1}^{N}$ without use of any entanglement resouces, e.g., two-particle
operations \cite{schliemann01}. Furthermore, any two $N$-particle Slater
determinants can be transformed to one another by means of a suitable unitary
operator $\hat{U}$ generated by a single-particle Hamiltonian.

The scattered state is obtained from a unitary evolution which involves
two-particle interactions and thus $|\tilde{f}\rangle$ in general is not
expected to be a simple Slater determinent, even if the incoming state is one.
Below, we analyze the complexity of $|\tilde{f}\rangle$ by testing its overlap
with any Slater determinant $|f_U\rangle \equiv \hat{U}|f\rangle$ which we
evolve from the initial state $|f\rangle$ via a unitary single-particle
operator $\hat{U}$ \cite{Vedral97}. Maximizing the overlap $\langle
f_U|\tilde{f}\rangle$ as a function of the evolution operator $\hat{U}$, we
find the optimal operator $\hat{U}_\mathrm{opt}$ providing the Slater
determinant $|f_{U_\mathrm{opt}} \rangle$ closest to $|\tilde{f} \rangle$. For
the chiral electrons discussed here, we find that the scattered state is again
a Slater determinant.

We restrict ourselves to physically relevant single-particle operators
$\hat{U}$ that can be generated through application of a local voltage-pulse
$V(t)$ at the position $x$ of the wire. The operator
\begin{equation} \label{eq:U_chi}
      \hat{U}_{\chi} = \hat{T}_+\exp\Bigl[ iv_{\rm \scriptscriptstyle F}\int dt\,
      \chi(t) \hat\rho(x,t) \Bigr]
\end{equation}
then adds the phase $\chi(t) = (e/\hbar)\int^t dt^\prime\, V(t^\prime)$
to each electron passing the position $x$. Furthermore, we change perspective
and rotate the scattered state $|\tilde{f}\rangle$ by $\hat{U}_{\chi}$.
In maximizing the overlap $\langle f| \hat{U}_{\chi}| \tilde{f}\rangle$,
one then has to find a pulse $V(t)$ such as to bring the resulting state as
close as possible to the original incoming state $|f\rangle$. The overlap
$\langle f|U_{\chi}| \tilde{f}\rangle$ can be expressed as
\begin{eqnarray}\label{eq:O}
      &&\langle f|U_{\chi}|\tilde{f}\rangle 
      = \int dx\,dx^\prime\, f^*_2(x) f_1(x^\prime)
      \nonumber\\
      &&\qquad \times \bigl\langle \hat{T}_K \bigl\{
      \hat{S}_K \hat{U}_{\chi K} \hat\Psi(x,t_2^-)
      \hat\Psi^\dagger(x^\prime,t_1^+)\bigr\} \bigr\rangle.
\end{eqnarray}
\\ The wave packets $f_{1,2}(x) = f(x-x_{1,2})$ are localized around the retarded
coordinates $x_{1,2}=v_{\rm\scriptscriptstyle F} t_{1,2}$ far left (right) to
the interaction region (thus rendering the overlap independent on the time
coordinates) and the unitary operator $\hat{U}_{\chi K}$ has a
nonvanishing phase $\chi(t)$ only on the upper branch of the Keldysh
contour (hence $\chi(t) = \chi(t^+)$). The average in Eq.\
(\ref{eq:O}) involves a product of exponentials linear in the bosonic fields;
in addition, we choose an incoming wave packet of Lorentzian form
\cite{ivanov97} $f_{\rm\scriptscriptstyle L}(x) = \sqrt{\xi/\pi}\,
(x+i\xi)^{-1}$ with width $\xi$. A straightforward calculation provides the
(zero-temperature) result
\begin{widetext}
\begin{equation}
   \langle f_{\rm\scriptscriptstyle L}|U_{\chi}
   |\tilde{f_{\rm\scriptscriptstyle L}}\rangle
   = \exp\Bigl[ -\int_0^\infty
   \frac{d\omega}{2\pi}\, \Bigr(v_{\rm\scriptscriptstyle F}^2|
   \chi(\omega)|^2 G_{++}(\omega)
   +2i\mbox{Re}[\chi(\omega)]
   e^{-\omega\tau_\xi}
   +\frac{i\omega_{\rm \scriptscriptstyle C}}2
   \frac{|g(\omega)|^2e^{-\omega\tau_\xi}}
   {1+i\omega_{\rm \scriptscriptstyle C}\,
   \Pi_{++}(\omega)/2} \Bigl( e^{-\omega\tau_\xi}
   -\frac{i\omega\chi(\omega)}{2\pi}\Bigr)\Bigr)\Bigr],
   \label{ov1}
\end{equation}
\end{widetext}
with the time-width $\tau_\xi=\xi/v_{\rm\scriptscriptstyle F}$ and the
transformed interaction kernel $g(\omega) = \int dt\,
g(v_{\rm\scriptscriptstyle F}t) e^{-i\omega t}$ (we assume $g(x)$ to be
centered around $x=0$).  $G_{++}(\omega)$ is the Fourier transform of the
bosonic Green's function $G_{++}(\omega,x,x) = \langle \hat{T}_+
\{\hat\rho(x,\tau) \hat\rho(x,0)\}\rangle$,
\begin{equation}
      G_{++}(\omega,x,x^\prime)
      = \frac1{2\pi v_{\rm\scriptscriptstyle F}^2} \int
      \frac{d\omega^\prime}{2\pi i}\,
      \frac{\omega^\prime\, e^{-i\omega^\prime(x-x^\prime)/
      v_{\rm\scriptscriptstyle F}}}{\omega^\prime
      -\omega-i\delta\,\mbox{sgn}(\omega)},
\end{equation}
and $\Pi_{++}(\omega)$ derives from $\Pi_{++}(\tau) = \langle \hat{T}_+
\{\hat{N}(\tau) \hat{N}(0)\}\rangle$, $\Pi_{++}(\omega)= \int dxdx^\prime
g(x)g(x^\prime) G_{++}(\omega,x,x^\prime)$.  Maximizing the overlap $\langle
f_{\rm\scriptscriptstyle L}|U_{\chi}|\tilde{f_{\rm\scriptscriptstyle
L}}\rangle$ with respect to the phase $\chi(t)$ one finds that the
optimal voltage-pulse must generate the phase
\begin{equation}\label{eq:phase_L}
      \chi_{\rm\scriptscriptstyle L} (\omega>0) = -\frac{\omega_{\rm
      \scriptscriptstyle C}}2\, \frac{|g(\omega)|^2 e^{-\omega\tau_\xi}}
      {1-i\omega_{\rm \scriptscriptstyle C}\,\Pi_{++}^*(\omega)/2}
\end{equation}
and $\chi_{\rm\scriptscriptstyle L}(\omega<0) =\chi_{\rm
\scriptscriptstyle L}^*(-\omega)$.

Calculating the overlap $\langle f_{\rm\scriptscriptstyle L}| U_{\chi_{\rm
\scriptscriptstyle L}}|\tilde{f}_{\rm\scriptscriptstyle L}\rangle$ for this
pulse one arrives at the result that the scattered state is a Slater
determinant state, $|\langle f_{\rm\scriptscriptstyle L}|U_{\chi_{\rm
\scriptscriptstyle L}}|\tilde{f}_{\rm\scriptscriptstyle L} \rangle| = 1$.
Hence, the (properly retarded) voltage-pulse generating the phase $\chi_{\rm
\scriptscriptstyle L}(t)$ rotates the scattered state $|\tilde{f}_{\rm
\scriptscriptstyle L}\rangle$ {\it exactly} back to the original state,
$\hat{U}_{\chi_{\rm\scriptscriptstyle L}} |\tilde{f}_{\rm \scriptscriptstyle
L}\rangle \propto |f_{\rm\scriptscriptstyle L}\rangle$, up to a phase factor.

The incoming state $|f_{\rm\scriptscriptstyle L}\rangle$ with a Lorentzian
wave packet above the Fermi sea can itself be obtained \cite{ivanov97} by
applying a voltage pulse of Lorentzian shape $V_{\rm\scriptscriptstyle L}(t) =
(2\hbar/e\tau_\xi)/ (1+(t/\tau_\xi)^2)$ with the associated phase $\phi_{\rm
\scriptscriptstyle L} (t) = 2\arctan(t/\tau_\xi)$, i.e., $|f_{\rm
\scriptscriptstyle L}\rangle = \hat{U}_{\phi_{\rm \scriptscriptstyle L}}
\hat{F}^\dagger |\Phi_{\rm \scriptscriptstyle F}\rangle$.  The operator
$\hat{U}_{\phi_{\rm \scriptscriptstyle L}}$ alone cannot add charge to the
system and it is the Klein factor that adds an additional electron at the
Fermi level.

Let us then apply an arbitrary voltage pulse $V(t)$ generating an
electron-hole state $|f_\phi\rangle = \hat{U}_\phi(\hat{F}^\dagger)^n
|\Phi_{\rm\scriptscriptstyle F} \rangle$ with $n$ excess electrons above the
Fermi sea.  Although not covering all possible incoming states, this procedure
generates an important family of states which can be {\it classically}
prepared by an experimenter. The action of the Coulomb interaction on this
incoming state can then be expressed through a change in the phase $\phi$,
i.e., the scattered state $\hat{S}|f_\phi\rangle$ is equivalent to the state
$\hat{U}_{\tilde\phi}(\hat{F}^\dagger)^n |\Phi_{\rm\scriptscriptstyle F}
\rangle$ with the phase
\begin{equation}\label{eq:mr}
   \tilde\phi(\omega>0) = \phi(\omega)\,
   \frac{1+i\omega_{\rm \scriptscriptstyle C}\,
   \Pi_{++}(\omega)/2}
   {1-i\omega_{\rm \scriptscriptstyle C}\,\Pi_{++}^*(\omega)/2}.
   \label{pulse}
\end{equation}
Note that it is the rotation by $\hat{U}_{\phi-\tilde\phi} \equiv
\hat{U}_{\chi} $ that plays the role of the optimal rotation Eq.\
(\ref{eq:U_chi}).

On a technical level, the reason for the separability of the scattered state
is found in the chiral nature of the scattering problem.  In the bosonic
language, the incoming state can be presented as a coherent state $|f_{\rm
\scriptscriptstyle L}\rangle = \prod_{k>0}e^{u_k\hat{b}_k - u_k^*
\hat{b}_k^\dagger}\hat{F}^\dagger|\mathrm{vac}\rangle$, with $u_k =
v_{\rm\scriptscriptstyle F}\sqrt{k}\, \phi_{\rm \scriptscriptstyle L}
(kv_{\rm\scriptscriptstyle F})\, e^{ikx}/2\pi$ and $|\mathrm{vac}\rangle$ the
bosonic vacuum. Since the bosonic Hamiltonian is quadratic and in the absence
of back reflection, the bosons only acquire a phase factor during the
propagation through the interaction region, $\hat{b}_k \rightarrow
e^{i\delta_k} \hat{b}_k$. Thus the scattered state is also a coherent state
with $u_k \rightarrow u_k e^{i\delta_k}$ and therefore exhibits the same
complexity as the incoming state.

\begin{figure} [t]
   \includegraphics[width=7.0cm]{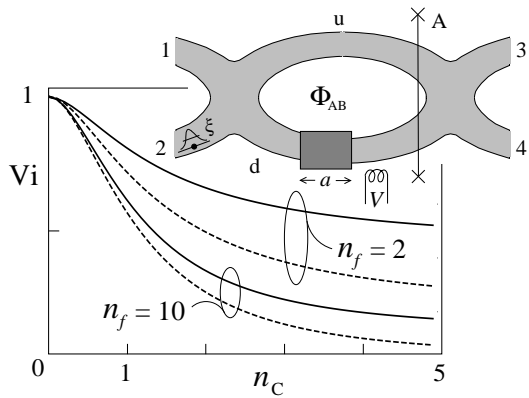}
   \caption[]
   {Top right: Asymmetric Mach-Zehnder interferometer with capacitive coupling
   to a metallic plate. The voltage pulse $V(t)$ serves to undo (part of) the
   shake-up due to the Coulomb interaction. Data: Corrected (solid line) and
   `bare' (dashed line) visibility as a function of the interaction strength
   $n_{\rm \scriptscriptstyle C} = \omega_{\rm \scriptscriptstyle C}
   \tau_a/4\pi$ for low- ($n_f = 2$) and high-energy ($n_f=10$) wave packets.}
   \label{fig:MZI}
\end{figure}

We now make use of the results Eqs.\ (\ref{eq:phase_L}) and (\ref{eq:mr}) in a
specific application, the improvement of the visibility in a Mach-Zehnder
interferometer (MZI) fed through single-electron pulses. The basic idea is to
undo the distortion of the wave packets by the Coulomb-interaction via
suitable voltage-pulses applied after the scattering region.  Consider an
asymmetric MZI where the electrons passing through one arm (the bottom arm d,
see Fig.\ \ref{fig:MZI}) are capacitively coupled to a metallic gate.  We
inject a single-electron wave packet $f(x)$ (at $x_0= v_{\rm
\scriptscriptstyle F}t_0<0$) through the lead 2 and calculate the average
excess current in lead 4 when a voltage-pulse with phase $\chi(t)$ is applied
right after the interaction region,
\[
      \frac{I_4(x_t)}{e v_{\rm\scriptscriptstyle F}} =
      (|t_\mathrm{u}|^2\! +|t_\mathrm{d}|^2)|f(x_t)|^2
      +t_\mathrm{u} t_\mathrm{d}^* f(x_t) {\cal G}_\chi(x_t)
      + c.c.,
\]
where $x_t=x\!-\!v_{\rm\scriptscriptstyle F}t$, $t_\mathrm{u(d)}$ are
amplitudes to go from lead $2$ to lead $4$ through the upper (lower) arm and
\begin{eqnarray}
      {\cal G}_\chi(x_t) &=& \int dx^\prime f^*(x^\prime-x_0) \label{G} \\
      &&\quad\times \bigl\langle \hat{T}_K \{ \hat{S}_K \hat{U}_{\chi K}
      \hat\Psi_\mathrm{d}(x^\prime,t_0^-)
      \hat\Psi_\mathrm{d}^\dagger(x,t^+)\} \bigr\rangle,
      \nonumber
\end{eqnarray}
with $\hat\Psi_\mathrm{d}$ the electronic field operator in the bottom arm of
the interferometer. The term $\propto {\cal G}_\chi(x_t)$ in $I_4$ describes the
oscillations in the average current and depends on the magnetic
flux $\Phi_{\rm\scriptscriptstyle AB}$ penetrating the interferometer.

The probability $P_4$ for the  excess electron to go into lead 4 can be
expressed as $P_4 = \int dt\, I_4(t)/e$. The visibility $\mathrm{Vi}$
is the amplitude of the flux-dependent oscillating contribution to $P_4$,
\begin{equation}
      \mathrm{Vi} = \Bigl| \int dx\, f(x) {\cal G}_\chi(x)\Bigr|.
      \label{V1}
\end{equation}
The voltage-pulse rotates the $\hat\Psi_\mathrm{d}(x,t)$ field in Eq.\
(\ref{G}), $\hat\Psi_\mathrm{d}(x,t) \rightarrow \hat{U}_{-\chi}
\hat\Psi_\mathrm{d} (x,t)\hat{U}_{\chi} = \hat\Psi_\mathrm{d}(x,t)
\exp[-i\chi(t-x/v_{\rm\scriptscriptstyle F})]$ and we can simplify
${\cal G}_\chi(x_t) = \exp[i\chi(-x_t/v_{\rm\scriptscriptstyle F})] \,
{\cal G}(x_t)$ with
\begin{eqnarray}\nonumber
   {\cal G}(x_t)
   \!&=&\!\! \int \! dx^\prime f^*(x^\prime-x_0) \langle
   \hat{T}_K\{ \hat{S}_K \hat\Psi_\mathrm{d} (x^\prime\!,t_0^-)
   \hat\Psi_\mathrm{d}^\dagger (x,t^+)\} \rangle \\
   \!&=&\!-\! \int\frac{dx^\prime}{2\pi i}\,
      \frac{f^*(x^\prime)}{x^\prime-x_t+i\delta}\,
      e^{i\Phi\bigl((x^\prime-x_t)/
      v_{\rm\scriptscriptstyle F}\bigr)}, \label{eq:G2} \\
   \Phi(t)
   \!&=&\! \frac{\omega_{\rm \scriptscriptstyle C}}2
      \int_0^\infty \frac{d\omega}{2\pi}
      \frac{|g(\omega)|^2}
      {1-i\omega_{\rm \scriptscriptstyle C}\,\Pi_{++}^*(\omega)/2}\,
      e^{i\omega t}.
      \label{Phi}
\end{eqnarray}
The optimal visibility $\mathrm{Vi}$ then is reached for a voltage-pulse
generating the phase $\chi_{\rm\scriptscriptstyle MZ}(t) = -\arg[f(-v_{\rm
\scriptscriptstyle F} t)]-\arg[{\cal G}(-v_{\rm\scriptscriptstyle F} t)]$. For
a Lorentzian wave packet $f_{\rm \scriptscriptstyle L}(x)$ with a width $\xi$
this optimized phase reduces to $\chi_{\rm\scriptscriptstyle MZ}(t) =
-\mbox{Re} \,\Phi(t+i\tau_\xi) = \chi_{\rm\scriptscriptstyle L}(t)/2$, half
the result Eq.~(\ref{eq:phase_L}).

In order to understand this result, we analyze the quantum state shared
between the two arms of the interferometer at point $A$, see Fig.\
\ref{fig:MZI}, $|A\rangle = t_\mathrm{2u}\,|f_{\rm \scriptscriptstyle
L}\rangle_\mathrm{u}\, \hat{U}_{\chi_{\rm\scriptscriptstyle MZ}} |\Phi_{\rm
\scriptscriptstyle F} \rangle_\mathrm{d} + t_\mathrm{2d}\,
|\Phi_{\rm\scriptscriptstyle F} \rangle_\mathrm{u}\, \hat{U}_{\chi_{\rm
\scriptscriptstyle MZ}} \hat{U}_{-\chi_{\rm \scriptscriptstyle L}} |f_{\rm
\scriptscriptstyle L}\rangle_\mathrm{d}$.  The visibility is $\mathrm{Vi}
\propto \bigl|\int dx\,\langle A|\hat\Psi_\mathrm{u}^\dagger(x,t)
\hat\Psi_\mathrm{d}(x,t)|A\rangle\bigr|$, or explicitly,
\begin{equation}
    \mathrm{Vi} = \big| _\mathrm{4} \langle \Phi_{\rm\scriptscriptstyle F}|
    \hat{U}^\dagger_{\chi_{\rm\scriptscriptstyle MZ}} 
    \hat\Psi_{f_{\rm \scriptscriptstyle L}} 
    \hat{U}_{\chi_{\rm\scriptscriptstyle MZ}} 
    \hat{U}_{-\chi_{\rm\scriptscriptstyle L}} 
    \hat\Psi_{f_{\rm \scriptscriptstyle L}}^\dagger 
    |\Phi_{\rm\scriptscriptstyle F} \rangle_\mathrm{4} \big| \label{V2}
\end{equation}
with $\hat\Psi_{f_{\rm \scriptscriptstyle L}}^\dagger = \int dx f_{\rm
\scriptscriptstyle L}(x) \hat\Psi^\dagger(x)$.  For a compensation $\chi_{\rm
\scriptscriptstyle MZ}$, the visibility is given by the overlap of the state
$\Psi_{f_{\rm \scriptscriptstyle L}}^\dagger \hat{U}_{\chi_{\rm
\scriptscriptstyle MZ}} |\Phi_{\rm\scriptscriptstyle F} \rangle$ (first term
in $|A\rangle$; the particle enters the lead 4 from the upper arm) and the
contribution $\hat{U}_{\chi_{\rm\scriptscriptstyle MZ}} \hat{U}_{-\chi_{\rm
\scriptscriptstyle L}}\Psi_{f_{\rm \scriptscriptstyle L}}^\dagger |\Phi_{\rm
\scriptscriptstyle F} \rangle$ (second term in $|A\rangle$ with the particle
coming from the lower arm). The different shapes of these wave functions
provide `which-path' information which implies a reduction in the visibility.
If the electron trajectory were known, one could fully compensate the effect
of the interaction by applying $\chi_{\rm \scriptscriptstyle MZ} =
\chi_{\rm\scriptscriptstyle L}$ every time when the electron chooses the lower
arm. Since this quantum information is not available, the pulse has to be
applied blindly and either compensates the effect of interaction (if the
particle indeed passed through the lower arm) or creates additional
electron-hole pairs (if the particle passed through the upper arm).  It turns
out, that a half-pulse $\chi_{\rm \scriptscriptstyle MZ} = \chi_{\rm
\scriptscriptstyle L}/2$ partly erases the information on the trajectory taken
by the electron: after recombination in the second beam splitter, the
many-particle states in the outgoing lead corresponding to the propagation
through the upper or the lower arm are optimally adjusted to each other such
as to increase the visibility. Or, in other words, the half-pulse
$\chi_{\rm\scriptscriptstyle L}/2$ on purpose erases part of the `which-path'
information in the outgoing lead.

In reality the form of the coordinate kernel $g(x)$ is not known. However, one
can measure the time dependence of the Aharonov-Bohm oscillations of the {\it
average} current and extract the phase $\Phi(i\tau_\xi-t)$.  Tracing the
maximum $I^>(t) = \mathrm{max}_{\Phi_{\rm\scriptscriptstyle AB}}
[I_4(t,\Phi_{\rm\scriptscriptstyle AB})]$ and minimum $I^<(t)$ currents as a
function of time for a Lorentzian wave packet, one can extract $\mbox{Im}
\Phi(i\tau_\xi-t)$ from the ratio $(I^>-I^<) /(I^>+I^<) = 2 \exp[\mbox{Im}
\Phi(i\tau_\xi-t)] \,|t_\mathrm{u}t_\mathrm{d}| /(|t_\mathrm{u}|^2 +
|t_\mathrm{d}|^2)$. The real part $\mbox{Re}\Phi(i\tau_\xi-t)$ follows from
the analytical properties of the function $\Phi(t)$.

We close with the explicit calculation of the visibility and its resurrection
for a model kernel $g(x) = \exp(-|x|/a)$ and an incoming Lorentzian. The
optimal voltage-pulse $V_{\rm\scriptscriptstyle L}(t) = e\,\partial_t
\chi{\rm\scriptscriptstyle MZ}(t)/\hbar$ derives from the phase 
\begin{equation}
    \chi_{\rm\scriptscriptstyle MZ}(t)
    =-4n_{\rm\scriptscriptstyle C} \int d\nu\,
    \frac{e^{i\nu t/\tau_a-|\nu|/n_f}}
    {(\nu+i)^2[(\nu-i)^2-n_{\rm\scriptscriptstyle C}]}.
\end{equation}

With $n=a\rho\,\delta\varepsilon$ the number of electrons in the interaction
region $a$ within the energy window $\delta\varepsilon$ ($\rho=1/\hbar
v_{\rm\scriptscriptstyle F}$ denotes the density of states), we can define the
two parameters  $n_f=a/\xi$ and $n_{\rm\scriptscriptstyle C}=\omega_{\rm
\scriptscriptstyle C}\tau_a/4\pi$ with the corresponding energy scales
$\delta\varepsilon_f = \hbar v_F/\xi$ associated with the excess energy
carried by the wave packet and $\delta\varepsilon_{\rm \scriptscriptstyle
C}=\hbar\omega_{\rm \scriptscriptstyle C}/2$ associated with the typical
energy of electron-hole excitations (electron-hole pairs with higher energies
are suppressed).
%
%
The (un)corrected visibilities ($\mathrm{Vi}$) $\mathrm{Vi}_\mathrm{opt}$
are given by
\begin{eqnarray}\label{eq:Vi}
   \mathrm{Vi} &=& \exp\Bigl[ 2in_{\rm \scriptscriptstyle C}
   \int d\nu \frac{\mbox{sgn}(\nu)\,
   e^{-2|\nu|/n_f}}{(\nu+i)^2[(\nu-i)^2-n_{\rm \scriptscriptstyle C}]}
   \Bigr], \\ \nonumber
   \mathrm{Vi}_\mathrm{opt}
   &=& \!\!\int\!\!\frac{d\kappa/\pi}{1+\kappa^2}
   \exp\Bigl[2in_{\rm \scriptscriptstyle C}
   \!\!\int\!\! d\nu\,\frac{\mbox{sgn}(\nu)\,
   e^{(i\kappa\nu-|\nu|)/n_f}} {(\nu\!+\!i)^2
   [(\nu\!-\!i)^2-n_{\rm \scriptscriptstyle C}]}\Bigr],
\end{eqnarray}
and are shown in Fig.~\ref{fig:MZI} as a function of interaction strength
$n_{\rm \scriptscriptstyle C}$ for high- and low-energy incoming wave packets.
The additional voltage-pulse indeed improves the visibility in both cases and
the correction is more efficient at large $n_{\rm \scriptscriptstyle C}$,
where the visibility saturates for fixed $n_f$. The saturation at large
Coulomb energy occurs due to energy conservation: an incoming electron cannot
excite electron-hole pairs with energy higher than $\delta \varepsilon_f$ no
matter how strong the Coulomb energy is. The enhancement in visibility is more
pronounced at large $n_{\rm \scriptscriptstyle C}$ and can reach of order
$100~\%$ of the original `bare' value.  Summarizing, we have established the
product nature of single-particle wave packets decohered through capacitive
Coulomb interaction and have demonstrated the possibility for their
resurrection through appropriate voltage pulses. In a real device, the
interaction may add dissipation to the system \cite{degiovanni09}, which our
scheme cannot cure.  Nevertheless, those parts of decoherence/dissipation
which are due to particle shake-up (i.e., a deformation of the wave-packet)
can always be compensated by application of proper voltage pulses.

We thank Gordey Lesovik for discussions and acknowledge the financial support
from the Swiss National Foundation through the Pauli Center at ETH Zurich.


\begin{thebibliography}{99}
\bibitem{norm} The functional integral over the $z(t)$ field is defined with
the norm, $\int D[z_\pm(t)] \equiv \prod_{t_i} \int dz_\pm(t_i)\,
[\omega_{\rm \scriptscriptstyle C}\,\delta t/(\pm\pi i) ]^{1/2}$
with $\delta t \rightarrow 0$.
\end{thebibliography}
\end{document}